# Calibrated Tree Priors for Relaxed Phylogenetics and Divergence Time Estimation

Joseph Heled[1] and Alexei J Drummond[1,2,3]

[1]Department of Computer Science, University of Auckland, Auckland, NZ
[2]Bioinformatics Institute, University of Auckland, Auckland, NZ
[3]Allan Wilson Centre for Molecular Ecology and Evolution, University of Auckland, Auckland, NZ

March 30, 2011

## Abstract

The use of fossil evidence to calibrate divergence time estimation has a long history. More recently Bayesian MCMC has become the dominant method of divergence time estimation and fossil evidence has been re-interpreted as the specification of prior distributions on the divergence times of calibration nodes. These so-called "soft calibrations" have become widely used but the statistical properties of calibrated tree priors in a Bayesian setting has not been carefully investigated. Here we clarify that calibration densities, such as those defined in BEAST 1.5, do not represent the marginal prior distribution of the calibration node. We illustrate this with a number of analytical results on small trees. We also describe an alternative construction for a calibrated Yule prior on trees that allows direct specification of the marginal prior distribution of the calibrated divergence time, with or without the restriction of monophyly. This method requires the computation of the Yule prior conditional on the height of the divergence being calibrated. Unfortunately, a practical solution for multiple calibrations remains elusive. Our results suggest that direct estimation of the prior induced by specifying multiple calibration densities should be a prerequisite of any divergence time dating analysis.

# Introduction

In addition to observed sequence data, a Bayesian phylogenetic analysis can incorporate other sources of knowledge through the application of informative priors. The use of so-called "soft calibrations" (Rannala and Yang, 2005), in the form of informative prior distributions on the divergence times of internal nodes during a phylogenetic analysis, have become increasingly common. This is especially the case in Bayesian phylogenetic models that support "relaxed phylogenetics", in which genetic distances are partitioned into divergence times and lineage-specific substitution rates using a relaxed molecular clock (Drummond et al., 2006).

Although these methods are now quite widely used, the statistical properties of prior distributions subject to calibration densities have not been carefully investigated, *when the ranked tree topology is a random variable*. In the relaxed phylogenetic models implemented in BEAST (Drummond and Rambaut, 2007), calibration is achieved by one of three means: (i) calibration of the rate of evolution through an informative prior on the substitution rate, (ii) calibration through heterochronous data (Drummond et al., 2002, 2003), (iii) calibration by specification of an informative prior distribution on the divergence time(s) of one or more internal nodes. Whereas the first two methods are relatively straightforward and have been well-studied, the statistical properties of the third option in a Bayesian setting have not been well studied when the tree is a random variable.

Here we aim to highlight some of the statistical properties of calibration on internal nodes in BEAST 1.5, a commonly used Bayesian MCMC implementation, and give a new method of constructing a calibration prior that has more intuitive statistical properties when the tree is a random variable. Below we give two examples illustrating how the current implementation of calibration in BEAST may induce non-uniform prior distribution on the tree topology, and how the marginal prior distribution of the calibrated nodes may differ from the calibration density used to construct the tree prior. While the form of the calibrated tree prior can be computed directly for simple cases, in general, the precise relationship between the calibration densities used to construct the tree prior and the actual marginal priors on the calibrated nodes can only be investigated by direct simulation of the tree prior using MCMC. When there is only a single calibration density, we introduce an alternative method of specifying the marginal prior distribution of the calibrated node directly. However when using the existing calibration method we recommend direction simulation of the tree prior as a standard pre-cursor to all relaxed phylogenetic analyses involving internal node calibration densities.

# The construction of a calibrated tree prior in BEAST

When calibrating the divergence times of some internal nodes, the tree prior is constructed in BEAST using three main ingredients:

1. One or more "calibration densities", each applied to the divergence time of the most recent common ancestor of a subset of the taxa.

2. A parametric "tree prior" and associated hyper-parameters and hyper-priors that specifies a density on the topology and all the divergence times of the tree.

3. Zero or more additional constraints on the topology in the form of subsets of taxa that are constrained to be monophyletic.

In BEAST, these ingredients are combined in a particular way to construct a prior distribution on time-trees. The combination of the latter two ingredients is quite unproblematic from the point of view of interpretation. The resulting distribution is simply the relevant parametric "tree prior" *conditional* on the topological constraints. Although this interpretation is simple, it is worth noting that the resulting distribution of both the divergence times and (obviously) the tree topology will differ from the unconstrained distribution.

However, the first of these ingredients can be incorporated into the model in a number of ways. A general method for computing a conditional birth-death-sampling prior for a tree with a fixed ranked topology has been described (Rannala and Yang, 2005), but this is not suitable when we wish to infer the topology. For the birth-death model, the special case where the node in question is the root is given in (Gernhard (2008) Theorem 4.1 with k=1). In BEAST, the calibration density is combined with the tree prior by simply taking their product. This ignores the overlap in state space of the two densities and we will call this the *multiplicative-construction*. In papers applying BEAST, the calibration density is often known as the "calibration prior" or the "prior on the divergence times", but we will avoid using the term prior, and use "calibration density" instead, since in the multiplicative-construction this distribution does not correspond to the marginal prior distribution of the associated divergence time. If the birth rate and the calibration density are really independent sources of information about the phylogeny then this may not be a bad method to construct the calibrated tree prior, although this construction certainly does not follow the rules of probability calculus. Specifically, the multiplicative-construction is problematic in situations where the researcher expects the calibration prior to represent the marginal distribution of the calibrated node, and can lead to unexpected results.

For example, consider associating a calibration density on $T_{AB}$, the time of the Most Recent Common Ancestor (MRCA) of A and B in a 3-taxon tree A, B, C. A Yule prior with a birth rate $\lambda = 1$ is used for the tree and an exponential density with mean 2 is used to calibrate $T_{AB}$. The density of $T_{AB}$ obtained by running BEAST using only the prior is shown in Figure 1(a). The same setup with a gamma ($\Gamma$) calibration density is shown in Figure 1(b).

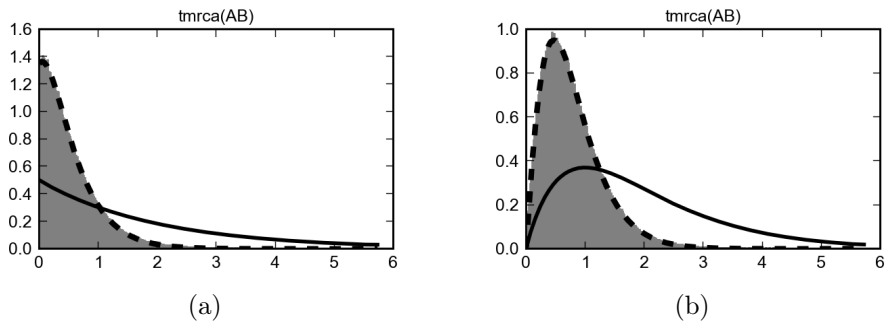

Figure 1: (a) Density of $T_{AB}$ from a BEAST run with a Yule prior ($\lambda = 1$) and exponential calibration density with mean 2 ($\mu = 0.5$). The calibration density is shown in black. The induced density matches the theoretical density (in dashed line).
(b) Similar to (a) but with a Gamma ($\Gamma(k = 2, \theta = 1)$) calibration density.

Note that those are the expected outputs. While difficult in general, here we can analytically compute the marginal prior density (shown as a dashed line), which exactly matches the distribution sampled by MCMC with BEAST. Also note that in case (a) the tree ((A, B), C) is preferred over the other two possible trees for any value of $\mu$. This may seem counter-intuitive at first because one might expect the pairing of A and B to be less probable when the mean of $T_{AB}$ is larger than the expected height of the tree. See Appendix A for more details and other examples.

**A multiple calibration data set of Marsupials**

As the number of calibration densities used to construct the tree prior grows it becomes hard to describe the joint prior on the calibrated node heights or their individual marginal priors analytically. However, it is always possible to examine the mismatch between the specified calibration densities and marginal densities that result from the multiplicative-construction. In Phillips et al. (2009) sequences of 7 nuclear genes and the complete mitochondrial (mt) genome protein-coding and RNA-coding DNA sequences for 7 placental mammals, 3 marsupials, 2 momotremes and 2 sauropsids were analyzed, aimed at dating the echidna-platypus divergence. Here we have re-run the MCMC analysis without the sequence data to show the marginal distributions that result from the multiplicative-construction (Fig. 2) on the eight calibrated nodes, alongside the calibration densities specified. We follow the authors by constraining mammals and sauropsids to be monophyletic. Since no specific prior was placed on the birth rate (implied improper prior between zero and infinity), it can be seen that the root height distribution almost matches the calibration density, but most of the others show strong modality, due to the interaction between the calibration densities and the topological constraints.

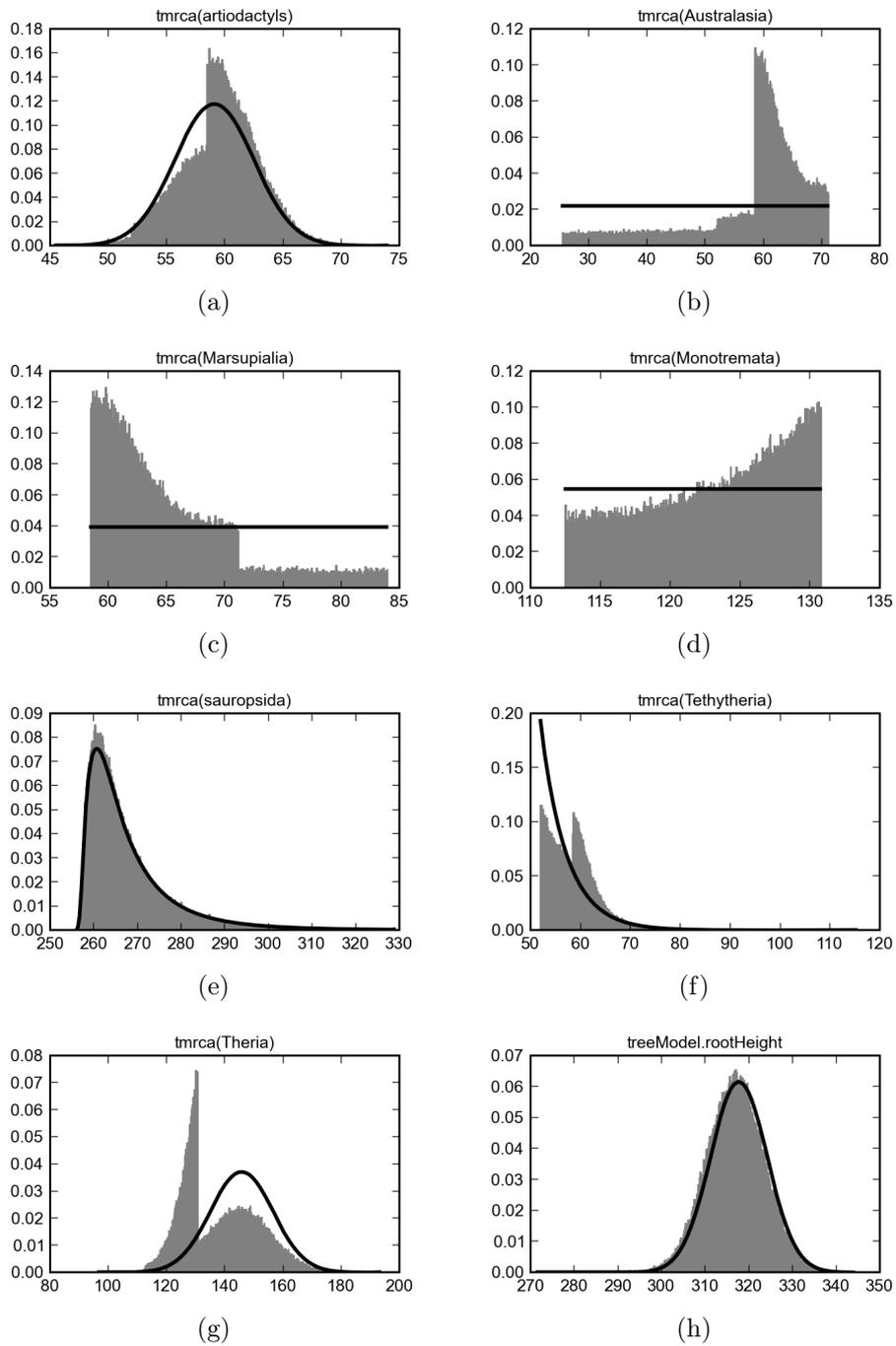

Figure 2: The marginal prior distributions that result from the multiplicative-construction (gray) versus calibration densities (black) specified for the calibrated nodes from Phillips et al. (2009). The marginal prior distributions were obtained from a MCMC run using the prior only. The calibration densities are as defined by the authors.

# Constructing a tree prior with an arbitrary marginal distribution on a $t_{MRCA}$

What are the desired properties of a calibrated tree prior? First, we would like the marginal density of the calibrated node to match the calibration density, and second, conditional on the calibrated node height, we want two trees to have relative prior densities proportional to some sensible generative process like the Yule, Birth-Death-Sampling (Stadler, 2009, 2010) or Coalescent (Kingman, 1982; Griffiths and Tavare, 1994) tree prior.

Let $\tau(g)$ be the TMRCA for calibrated taxa on genealogy $g$ from the space of all genealogies $G$. Consider the function $\rho_\text{G}(g)$, a candidate for a calibrated tree prior density on the space of genealogies and $\rho_\text{T}(\cdot)$, the desired marginal calibration density on $\tau(g)$. The following properties are desired:

(I) The marginal density on the calibrated node is equal to the calibration density:

$$\rho_\text{T}(x) = \int_{\substack{g \in G \\ \tau(g)=x}} \rho_\text{G}(g) = \int_{g \in G} 1(\tau(g) = x)\rho_\text{G}(g) \mathrm{d}g \tag{1}$$

In words, the total density of all trees for which the TMRCA of calibrated taxa is $x$ equals the calibration density at $x$. The integral is written first informally, then formally using the indicator function $1(\cdot)$, which is equal to 1 when the argument is true, 0 otherwise. Note that this is an integral over all genealogies, or time-trees. When the genealogy is represented as $g = \{\psi, h\}$, where $\psi$ is the ranked topology and $h$ is a vector of internal node heights in order, the integral can be written as

$$\rho_\text{T}(x) = \sum_{\psi \in \Psi} \int 1(\tau(\{\psi, h\}) = x)\rho_\text{G}(\{\psi, h\}) \, \mathrm{d}h. \tag{2}$$

(II) When restricted to a subset of trees with equal calibrated node height, the density is proportional to $f_G$ density:

$$\tau(g_1) = \tau(g_2) \implies \frac{\rho_\text{G}(g_1)}{\rho_\text{G}(g_2)} = \frac{f_\text{G}(g_1)}{f_\text{G}(g_2)}. \tag{3}$$

Constructing $\rho_G : G \to \mathbb{R}$ which satisfies (I) and (II) is quite easy:

$$\rho_G(g) = f_G(g) \frac{\rho_T(\tau(g))}{f_T(\tau(g))}, \qquad (4)$$

where $f_T(\cdot)$ is the marginal distribution of $\tau$ under $f_G$. We call this the *conditional-construction*. Informally, equation 4 can be written as

$$\text{new-joint-prior} = \text{old-joint-prior} \times \frac{\text{new-marginal}}{\text{old-marginal}}.$$

It is easy to see that $\rho_G(\cdot)$ satisfies (I) and (II), and integrates to 1. For genealogies with equal calibration the calibration and marginal are equal, so their ratio is $f_G$ as the other two terms cancel (II). And when integrating over trees with equal calibration, the calibration and marginal can be moved out of the integral, which leaves only the $f_G$ term inside, which then cancels with the marginal, leaving the calibration (I).

The conditional-construction is useful in practice only if the marginal density $f_T(\cdot)$ can be computed efficiently

$$f_T(x) = \int_{g \in G} 1(\tau(g) = x) f_G(g) \mathrm{d}g, \qquad (5)$$

Note that the domain of $f_G(\cdot)$ may depend on the conditions imposed on $g$. If taxa $\phi$ if not required to be monophyletic, the domain is all genealogies ($G$). When $\phi$ is required to be monophyletic, the domain is all genealogies which have $\phi$ as a monophyletic clade ($G_\phi$).

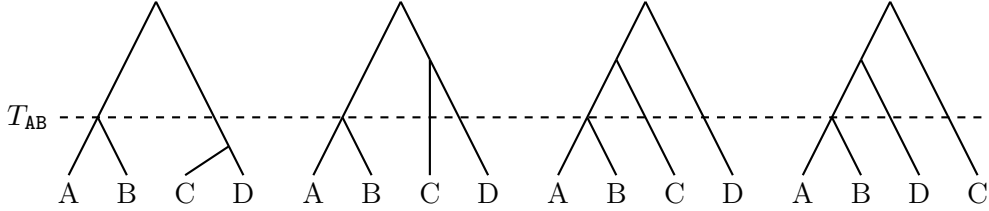

Figure 3: The four ranked trees with 4 taxa, (A,B) monophyletic showing that all have equivalent $\tau_{AB}$

## Yule tree prior on four taxa with one monophyletic calibration

We now describe in detail how to compute the marginal prior (equation 5) for the Yule tree prior with calibration on a 2-taxa monophyletic clade `(A,B)` in a 4-taxa tree. We then show how the same can be done in the general case for Yule tree prior on $n$ taxa and one calibration.

There are four ranked trees (Fig. 3). One, in which $T_{CD}$ is lower than $T_{AB}$ (case 1), and 3 ranked trees where $T_{AB}$ has the most recent divergence time (case 2).

Let $T = (T_1, T_2, T_3)$ be the intra-coalescent time intervals. For example in case 1 the interval between the leaves and $T_3 = T_{CD}$, $T_2 = T_{AB} - T_{CD}$ and so on. Under the Yule prior each interval is distributed exponentially, $T_{i-1} \sim \text{Exp}(i\lambda)$, and the joint density for $T$ is

$$f_Y(t) = 24\lambda^3 e^{-\lambda(4t_3 + 3t_2 + 2t_1)}$$

Since for case 1 we have $T_{AB} = T_2 + T_3$, the marginal density is given by

$$u^{(1)}(t_{AB}) = \int_0^\infty \int_0^{t_{AB}} f_Y(t_1, t_{AB} - t_3, t_3) \, dt_3 \, dt_1$$

$$= \int_0^\infty \int_0^{t_{AB}} 24\lambda^3 e^{-\lambda(4t_3 + 3(t_{AB} - t_3) + 2t_1)} \, dt_3 \, dt_1$$

$$= 12\lambda(e^{-3\lambda t_{AB}} - e^{-4\lambda t_{AB}}) \qquad (6)$$

Note the range $[0, t_{AB}]$ in the integral of $t_3$, which keeps the branch length positive. For case 2, we obtain

$$u^{(2)}(t_{AB}) = \int_0^\infty \int_0^\infty f_Y(t_1, t_2, t_{AB}) \, dt_2 \, dt_1 \quad = \quad 4\lambda e^{-4\lambda t_{AB}} \qquad (7)$$

Since there are three ranked trees with density $u^{(2)}$ and one with $u^{(1)}$, the marginal Yule distribution is given by

$$f(t_{\text{AB}}) = \frac{1}{4}u^{(1)}(t_{\text{AB}}) + \frac{3}{4}u^{(2)}(t_{\text{AB}}) = 3\lambda e^{-3\lambda t_{\text{AB}}}. \tag{8}$$

**Yule tree on four taxa with one calibration prior, no monophyly**
The construction for the monophyletic clade can be adapted to placing a calibration on $T_{\text{AB}}$ without enforcing monophyly. Instead of two cases we have three: A, B is monophyletic (case I), the common ancestor of A, B has 3 descendants (case II), and the common ancestor of A, B is the root (case III). We already have the densities when A, B is monophyletic, and the density for case II is given by equation (6). We still need a density for case III:

$$u^{(3)}(t_{\text{AB}}) = \int_0^{t_{\text{AB}}} \int_0^{t_{\text{AB}}-t_3} f_Y(t_{\text{AB}} - t_3 - t_2, t_2, t_3) \, dt_2 \, dt_3$$
$$= 12\lambda e^{-2\lambda t_{\text{AB}}}(1 - e^{-\lambda t_{\text{AB}}})^2.$$

The three densities are combined by weighting them according to the number of ranked topologies to which they apply. For case I we have, as before, 1 and 3 ranked topologies with densities $u^{(1)}$ and $u^{(2)}$. For case II there are 4 ranked topologies with density $u^{(1)}$, and for case III there are 10 with density $u^{(3)}$. Together we get

$$f(t_{\text{AB}}) = \frac{10u^{(3)} + 3u^{(2)} + 5u^{(1)}}{10 + 3 + 5} = \frac{\lambda e^{-2\lambda t_{\text{AB}}}}{3}(12e^{-2\lambda t_{\text{AB}}} - 30e^{-\lambda t_{\text{AB}}} + 20).$$

## Yule tree prior with one monophyletic calibration prior

The four taxa case can be generalized to any monophyletic clade $\phi$ of size $n_c$ in an $n = n_c + n_o$ taxa tree. Formally, the genealogy $g$ is a pair $g = \{\psi, h\}$, where $\psi$ is the ranked topology and $h$ is a vector of the internal node heights in reverse order, $h = (h_1, h_2, \cdots, h_{n-1})$. Since $\phi$ is monophyletic it is one of the ranked topologies in $\Psi_\phi$, the set of ranked topologies containing that clade.

Now, the Yule density for the heights is equally divided between all ranked trees having those heights. Since there are $|\Psi_\phi|$ of them, the density for the genealogy $g$ is

$$f_Y(g) = \frac{1}{|\Psi_\phi|} n! e^{-\lambda h_1} \prod_{i=1}^{n-1} \lambda e^{-\lambda h_i}. \tag{9}$$

Define $i(\psi)$ as the rank of the MRCA of $\phi$. The marginal Yule density is given by

$$f_{T_\phi}(x) = \sum_{\psi \in \Psi_\phi} \int_{h_1=x}^{\infty} \cdots \int_{h_{i(\psi)-1}=x}^{h_{i(\psi)-2}} \int_{h_{i(\psi)+1}=0}^{x} \cdots \int_{h_{n-2}=0}^{h_{n-3}} \int_{h_{n-1}=0}^{h_{n-2}} f_Y(\{\psi, h\})$$

Surprisingly, this multiple integral evaluates to a simple expression which depends only on the size of $\phi$ and does not depend on $n$ (Appendix C)

$$f_{T_\phi}(x) = \frac{n_c^3 - n_c}{2} \lambda e^{-3\lambda x}(1 - e^{-\lambda x})^{n_c - 2}. \tag{10}$$

**Yule tree prior with one calibration prior, no monophyly**

Deriving the conditional density for the age of the most recent common ancestor of a subset of taxa $\phi$, without the constraint of monophyly of $\phi$ is more involved. Again assume a subset $\phi$ of size $n_c$ in an $n = n_c + n_o$ taxa tree. The conditional density is broken into $n_o + 1$ cases:

**Case 0:** taxa set $\phi$ is monophyletic,

**Case 1:** the most recent common ancestor of taxa $\phi$ has $n_c + 1$ descendants,

**Case 2:** the most recent common ancestor of taxa $\phi$ has $n_c + 2$ descendants,

⋮

**Case $n_o$:** the most recent common ancestor of taxa $\phi$ is the root.

Since in case $k$ we have a monophyletic clade of size $k + n_c$, the density for that case is given by $f_{T_{\phi_k}}(x)$ (equation 10), where $\phi_k$ is of size $k + n_c$. Note that in equation (10) only the size of the clade matters. To combine the densities from all cases into an overall density we need the number of ranked topologies for each case. Those counts, when scaled to add to 1, act as the coefficients $w_k$ in the final equation.

$$\tilde{f}_{T_\phi}(x) = \sum_{k=0}^{n_o} w_k f_{T_{\phi_k}}(x). \tag{11}$$

For the derivation of $w_k$ see Equation (13) in Appendix B.

Note that the formula works for the special case $n_c = 1$, that is when we wish to condition on the time a particular taxon "attaches" to the tree. In that case the marginal density has the simple form $2\lambda e^{-2\lambda x}$.

# Revisiting an analysis of the Chacma Baboon (*Papio ursinus*)

Sithaldeen et al. (2009) provide a phylogenetic analysis of the Chacma Baboon sequences sampled across the entire range of the species. The authors analyze 52 mtDNA samples, using a Yule prior coupled with calibration densities on the root and two nested monophyletic clades. While this calibrated tree prior has multiple calibration densities and therefore does not fall under the cases previously described, we can derive the marginal density in this particular case using the same methods (Appendix D).

Before applying the new prior we run both the original analysis and a prior-only version. Figure 4 shows the calibration density as specified in the BEAST analysis for the 3 nodes, together with the induced density from the prior-only run and the posterior values from the full analysis. We can clearly see that the posterior values match the induced prior almost to perfection, and that the induced prior is shifted in varying degrees from the calibration priors due to the interaction with the Yule prior. It is not really surprising that the analysis was able to "match" all three marginal divergence time priors since it used a relaxed clock with a wide and flat prior on the rate mean and variance, accommodating branch rate/time combinations whose products satisfy the desired branch length in substitutions while also producing branch times that closely match the marginal tree prior on the calibrated nodes.

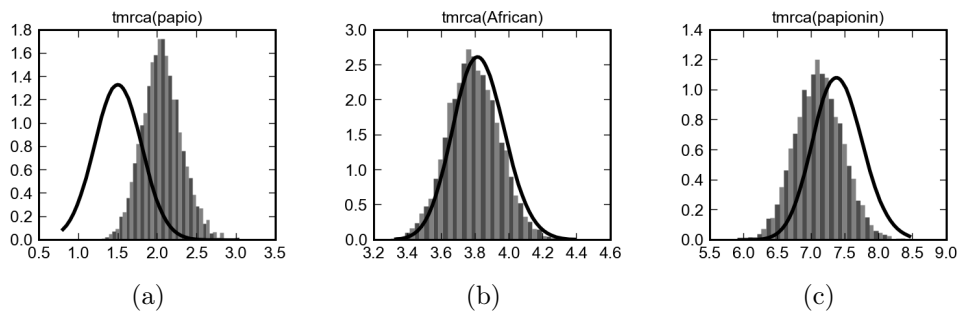

(a)      (b)      (c)

Figure 4: Induced marginal prior distributions (gray) versus calibration densities (black) versus marginal posterior distributions (dark gray) for the calibrated nodes from (Sithaldeen et al., 2009). The induced marginal prior distributions were obtained from a MCMC run using the prior only. The calibration densities are as defined by the authors.

Given that the data had no visible effect on the posterior distribution of the calibrated divergence times, it is reasonable to assume the prior plays a significant role in the divergence times of the non-calibrated nodes as well. This is indeed the case. Figure 5 shows the trees with the key divergence times from the original, independent construction run (a), and the tree from the run with the conditional construction (b). We can see that the lowest calibration node in (b) matches the expected mean of the calibration prior, and as a result all the divergence times in the subtree below have earlier times as well.

# Discussion and Conclusions

It is sometimes possible to construct a calibrated tree prior that factorizes precisely into a standard process-based tree prior conditional on the divergence times of the calibrated nodes and an independent marginal prior on those divergence times. We have demonstrated this for one calibrated node in the Yule prior. In order to produce such a conditional-construction, one simply needs to be able to efficiently calculate the marginal distribution of the calibrated nodes under the uncalibrated tree prior of choice.

Other conditions on tree priors are also possible. For example, conditioning on the root height of the tree is fairly straightforward for both the Yule model and the more general birth-death model of speciation (Stadler, *pers. comm.*). In fact, the original description of both the Yule and birth-death models in a phylogenetic context were in the form of a conditioning on the root height. However, those formulations did not condition on the number of taxa, which is also required. Nevertheless, arriving at a Yule probability density conditional on both the root height and the number of taxa is straightforward from that earlier work.

We are fairly confident that the methods presented here can be extended to handle multiple marginal prior distributions on internal nodes. However, the formulas for the conditional densities grow rapidly in size as a function of the number of conditions and taxa. As a result, the determination and evaluation of those conditional priors may become a practical problem.

The method that BEAST implements for constructing calibrated tree priors can lead to marginal distributions on calibrated nodes which are very different than the calibration densities chosen, as seen in Figure 2. In practice, any multiple-calibration analysis should always involve direct computation of the calibrated tree prior (by MCMC), and preferably report the actual marginal calibration prior for nodes of interest. Finally, in general, both multiplicative-construction and the conditional-construction produce non-uniform distributions on the (ranked) topology.

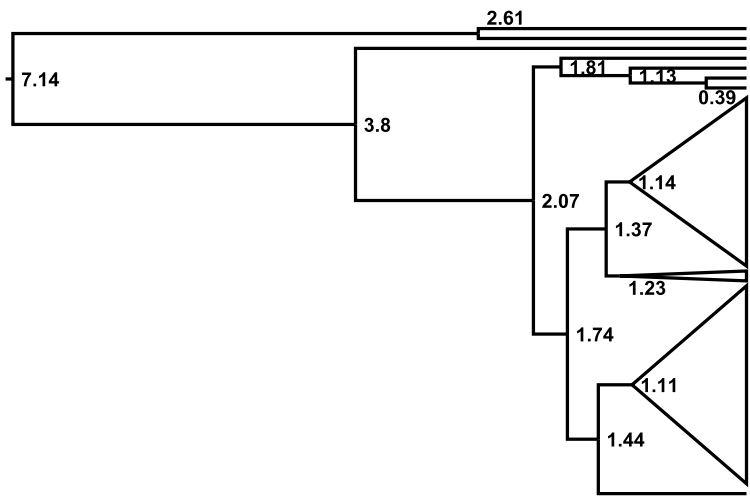

(a)

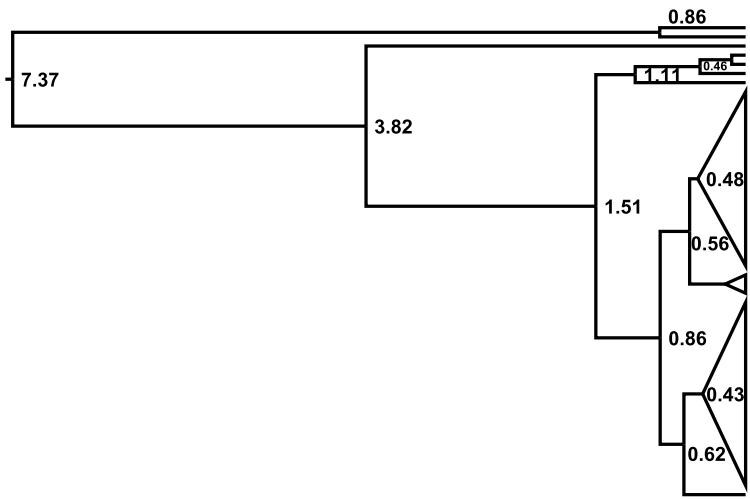

(b)

Figure 5: (a) Posterior estimates for divergence times using the independent construction prior. (b) Posterior estimates for divergence times using the conditional construction prior.

# Acknowledgements

We would like to thank David Bryant, Ron DeBry, Laura Kubatko and an anonymous reviewer for helpful comments. This work was supported by Marsden grant #UOA0502.

# A Examples of calibrated tree priors using the multiplicative-construction

In the multiplicative-construction used by BEAST 1.5, the tree topology and divergence times are influenced both by the calibration density and by the birth rate ($\lambda$) of the Yule model of tree branching. These two sources of information are combined to construct a prior density on the tree.

For our first example we consider associating a calibration density with $T_{\texttt{AB}}$, the time of the Most Recent Common Ancestor (MRCA) of `A` and `B` in a 3-taxon tree `A,B,C`. A Yule prior with a birth rate $\lambda$ is used for the tree and an exponential density with mean $1/\mu$ is used to calibrate $T_{\texttt{AB}}$. Label with $T_2$ the time of the youngest internal node and with $T_1$ the time between the root and the youngest internal node. Under the Yule model $T_1 \sim \text{Exp}(2\lambda)$, $T_2 \sim \text{Exp}(3\lambda)$ and the calibration density is $T_{\texttt{AB}} \sim \text{Exp}(\mu)$. Obviously, since $T_{\texttt{AB}}$ is a function of $T_1, T_2$, the Yule model already implies a marginal density on $T_{\texttt{AB}}$, so the Yule density and the calibration density share state space and are not independent.

In BEAST 1.5, this is ignored and the densities are simply multiplied together to form the multiplicative-construction. The resulting expression is not a proper density, but it is trivial to normalize by constant factor $1/z$. The resulting probabilities of the three possible topologies are,

$$f(\texttt{(A,B),C}) = \frac{1}{z}\int_0^\infty \int_0^\infty 2\lambda e^{-2\lambda t_1} 3\lambda e^{-3\lambda t_2} \mu e^{-\mu t_2} \, \mathrm{d}t_1 \mathrm{d}t_2 = \frac{1}{Z}\frac{3\lambda\mu}{3\lambda+\mu}$$

$$f(\texttt{(A,C),B}) = f(\texttt{(B,C),A})$$

$$= \frac{1}{z}\int_0^\infty \int_0^\infty 2\lambda e^{-2\lambda t_1} 3\lambda e^{-3\lambda t_2} \mu e^{-\mu(t_1+t_2)} \, \mathrm{d}t_1 \mathrm{d}t_2$$

$$= \frac{1}{z}\frac{6\lambda^2\mu}{(3\lambda+\mu)(2\lambda+\mu)}$$

The constant factor $z$ is easily computed, since the three integrals have to sum to 1, giving $z = \frac{3\lambda\mu(6\lambda+\mu)}{(2\lambda+\mu)(3\lambda+\mu)}$. So, the relative ratio of the three topologies is $2\lambda + \mu : 2\lambda : 2\lambda$ and tree `((A,B),C)` is preferred for any value of $\mu$. Furthermore it can be shown that under the multiplicative-construction $E[T_2] = 1/_{\mu+3\lambda}$ instead of $1/_{3\lambda}$ under the Yule, while $E[T_1] = 1/_{2\lambda}(1 - {4\mu\lambda}/{(\mu+2\lambda)(\mu+6\lambda)})$ instead of $1/_{2\lambda}$.

One may think that placing a calibration density on the non-monophyletic clade is the cause of the problem. However, we can repeat the calculation for a 4-taxon tree while enforcing monophyly of A,B. In one of the four possible topologies (first left in figure 3), the TMRCA of (A,B) is the larger of the two internal nodes ($T_{\texttt{AB}} = T_2 + T_3$), and is the smaller of the two in the other 3 cases ($T_{\texttt{AB}} = T_3$). The total densities for those two cases are

$$f_3(t_{AB}) = \int_0^\infty \int_0^\infty \int_0^\infty 2\lambda e^{-2\lambda t_1} 3\lambda e^{-3\lambda t_2} \lambda e^{-4\lambda t_3} \mu e^{-\mu t_3} \, \mathrm{d}t_1 \mathrm{d}t_2 \mathrm{d}t_3$$

$$= \frac{4\lambda\mu}{\mu + 4\lambda}$$

$$f_{23}(t_{AB}) = \int_0^\infty \int_0^\infty \int_0^\infty 2\lambda e^{-2\lambda t_1} 3\lambda e^{-3\lambda t_2} \lambda e^{-4\lambda t_3} \mu e^{-\mu(t_2+t_3)} \, \mathrm{d}t_1 \mathrm{d}t_2 \mathrm{d}t_3$$

$$= \frac{12\lambda^2 \mu}{(\mu + 3\lambda)(\mu + 4\lambda)}.$$

Now, since there are two ranked topologies with the unranked topology ((A,B),(C,D)), the ratio is

$$f(((\texttt{A,B}),\texttt{C}),\texttt{D})) : f(((\texttt{A,B}),(\texttt{C,D}))) = \frac{f_3(t_{AB})}{f_3(T_{AB}) + f_{23}(T_{AB})}$$

$$= \frac{\mu + 3\lambda}{\mu + 6\lambda}.$$

So, a ratio of $\mu + 6\lambda : \mu + 3\lambda : \mu + 3\lambda$ is obtained for the 3 topologies ((A,B),(C,D)), (((A,B),C),D) and (((A,B),D),C). Again, the first topology is preferred regardless of $\mu$.

Even when restricting the 3 taxon tree to a single topology by enforcing monophyly, the induced prior on divergence times depends on the specific interaction between the tree prior and the calibration density. Consider a Yule prior with birth rate $\lambda$ and a gamma prior with shape 2 and scale $\theta$ ($\Gamma(2,\theta)$) on $T_{\texttt{AB}}$. The expected divergence time under this combination can be shown to be $\frac{2\theta}{1+3\lambda\theta}$; which would always be less than the mean of the calibration density, $2\theta$. Finally, instead of fixing $\lambda$ we can assume a hyperprior on $\lambda$ - a very common practice in BEAST. This results in an increase in the dimensionality of the state space, and when deriving expectations or clade probabilities we need to integrate over the divergence times and $\lambda$ to obtain the constant normalizing factor. To compute the expected divergence time of (A,B) where $\lambda$ has a uniform hyperprior of $[0, N]$, we first derive the constant normalization factor,

$$z = \int\limits_0^N \int\limits_0^\infty \int\limits_0^\infty 2\lambda e^{-2\lambda t_1} 3\lambda e^{-3\lambda t_2} \frac{t_2 e^{-t_2/\theta}}{\theta^2} \frac{1}{N} \mathrm{d}t_1 \mathrm{d}t_2 \mathrm{d}\lambda$$

$$= \int\limits_0^N \frac{3\lambda}{(3\lambda\theta + 1)^2 N} \mathrm{d}\lambda$$

$$= \frac{(3N\theta + 1)\log(3N\theta + 1) - 3N\theta}{3(3N\theta + 1)N\theta^2}.$$

The expectation under the multiplicative-construction is

$$E[T_2] = \frac{1}{z} \int\limits_0^N \int\limits_0^\infty \int\limits_0^\infty t_2 \, 2\lambda e^{-2\lambda t_1} 3\lambda e^{-3\lambda t_2} \frac{t_2 e^{-t_2/\theta}}{\theta^2} \frac{1}{N} \mathrm{d}t_1 \mathrm{d}t_2 \mathrm{d}\lambda$$

$$= \frac{1}{z} \int\limits_0^N \frac{6\lambda\theta}{(3\lambda\theta + 1)^3 N} \mathrm{d}\lambda$$

$$= \frac{9 N^2 \theta^3}{(3N\theta + 1)((3N\theta + 1)\log(3N\theta + 1) - 3N\theta)}$$

$$\approx \frac{\theta}{\log(3N\theta + 1) - 1}, \text{ for large } N.$$

with $N = 100$, the average divergence time is approximately $\frac{\theta}{\log(300\theta+1)-1}$, which is less than $\theta$ for any $\theta > 0.006$.

# B  Number of ranked topologies for a non-monophyletic clade

Here we derive the coefficients $w_k$ used in the formula for calculating the conditional density for the time of the MRCA of the non-monophyletic taxa set $\phi$. The coefficient $w_k$ is the ratio of $r_k$, the number of ranked topologies for case k, to the total number $\mathcal{R}_n$ of ranked topologies for an $n = n_c + n_o$ taxa tree.

$$w_k = \frac{r_k}{\mathcal{R}_n} \qquad \left(\sum_k w_k = 1\right)$$

$$\mathcal{R}_n = \prod_{i=2}^{n} \binom{i}{2}$$

Here, $r_k$ is the number of ranked topologies where $n_c$ taxa are part of a $n_c + k$ taxa sub-tree. The number is the product of

(i) the number of ways to choose $k$ taxa from $n_o$ to be part of the clade $\phi$

(ii) $\mathcal{C}_{k,n_c}$, the number of ranked trees with $n_c + k$ taxa whose common ancestor of $\phi$ is the root

(iii) $\mathcal{D}_{n_o-k,n_c+k}$, the number of ways to combine the remaining $n_o - k$ taxa with the clade in (ii).

(i) is simply $\binom{n_o}{k}$. For (ii) we start with the $\mathcal{R}_{n_c}$ ways to coalesce $n_c$ lineages. For each of those we can add the remaining $k$ in some fixed order. The first lineage can attach itself to $2+3+\cdots+n_c = \binom{n_c+1}{2}-1$ places to create a different ranked topology, the second to $2+3+\cdots+n_c+(n_c+1) = \binom{n_c+2}{2}-1$, and so on, giving

$$\mathcal{C}_{k,n_c} = \mathcal{R}_{n_c} \prod_{i=n_c+1}^{k+n_c} \left[\binom{i}{2} - 1\right]$$

Let $\mathcal{D}_{l,m}$ be the number of ways to combine $l$ lineages and a fixed sub-tree with $m$ lineages. Examine the possible choices for the first coalescent: Either two of the $l$ lineages are joined ($\binom{l}{2}$ ways), or this is a coalescent in the sub-tree. This observation leads to the following recursive formula

$$\mathcal{D}_{l,m} = \mathcal{D}_{l,m-1} + \binom{l}{2}\mathcal{D}_{l-1,m}.$$

With the initial condition $\mathcal{D}_{0,m} = 1$ and $\mathcal{D}_{l,1} = \mathcal{R}_{l+1}$. It is easy to show the above has the solution

$$\mathcal{D}_{l,m} = \binom{l+m}{l-1}\mathcal{R}_l. \tag{12}$$

Substituting $n_o - k$ for $l$ and $n_c + k$ for $m$ gives the required count for (iii). All three put together give

$$r_k = \binom{n_o}{k} \prod_{i=3}^{n_c} \binom{i}{2} \prod_{i=n_c+1}^{k+n_c} \left[\binom{i}{2} - 1\right] \prod_{i=2}^{n_o+1-k} \frac{i(i+n_c+k)}{2}. \tag{13}$$

## B.1 Computational note

The counts $r_k$ are large and we need to evaluate $w_k$ directly. Some tedious manipulations result in an expression which does not involve large numbers

$$w_k = \binom{n_o + n_c}{2}^{-1} \prod_{i=0}^{k-1} \frac{n_o - i}{i+1}\left(1 - \frac{2}{(i+n_c)(i+n_c+1)}\right)$$
$$\prod_{i=k}^{n_o-2} \frac{(i-k+2)(1+\frac{2}{i+n_c})}{i+n_c+1}.$$

# C Conditional density of the TMRCA for a monophyletic clade $\phi$

Here we derive the simple form of the marginal Yule density when the genealogy has a single monophyletic clade $\phi$ of size $n_c$ in a tree with $n$ taxa.

First note that the total number of those genealogies can be obtained from equation (12)
$$|\Psi_\phi| = \mathcal{R}_n^{(n_c)} = \mathcal{R}_{n_c} \mathcal{D}_{n-n_c,n_c}.$$

Partition all ranked topologies according to $i_\phi(\psi) = k + 1$, that is group together topologies having $k$ heights above the root of $\phi$.

$$f_{T_\phi}(x) =$$
$$\sum_{k=1}^{n_o} \sum_{\substack{\psi \in \Psi_\phi \\ i_\phi(\psi)=k}} \int_{h_1=x}^{\infty} \cdots \int_{h_{k-1}=x}^{h_{k-2}} \int_{h_{k+1}=0}^{x} \cdots \int_{h_{n-2}=0}^{h_{n-3}} \int_{h_{n-1}=0}^{h_{n-2}} f_Y(\{\psi, h\})$$

Under both conditions the multi-integral has the same value in each case. The integrals can be separated into two independent groups, the $n - k - 2$ heights below $x$ ($n_c - 2$ from $\phi$, $n_o - k$ from outside), and the $k$ heights above $x$. The first group integrates to $\frac{1}{(n-k-2)!}(1 - e^{-\lambda x})^{n-k-2}$, the second to $\frac{1}{(k+1)!}(e^{-\lambda x})^{k+1}$. Both from the simple observation that the integral on $k$ unrestricted heights is $k!$ times the integral on the order statistic. The root of $\phi$ contributes $\lambda e^{-\lambda x}$, giving

$$f_{T_\phi}(x) = $$
$$\lambda \sum_{k=1}^{n_o} \sum_{\substack{\psi \in \Psi_\phi \\ i_\phi(\psi)=k}} \frac{1}{\mathcal{R}_n^{(n_c)}} \frac{n!}{(n-k-2)!(k+1)!} (1 - e^{-\lambda x})^{n-(k+2)} (e^{-\lambda x})^{k+2}$$
$$= \lambda \sum_{k=1}^{n_o} \left( \frac{1}{\mathcal{R}_n^{(n_c)}} \frac{n!}{(n-k-2)!(k+1)!} (1 - e^{-\lambda x})^{n-(k+2)} (e^{-\lambda x})^{k+2} \right) \sum_{\substack{\psi \in \Psi_\phi \\ i_\phi(\psi)=k}} 1$$

The last step is possible because none of the terms depend on the specific topology. The number of ranked topologies under our criteria is,

$$\sum_{\substack{\psi \in \Psi_\phi \\ i_\phi(\psi)=k}} 1 = \binom{n-k-2}{n_c-2} \mathcal{R}_{k+1} \mathcal{R}_{n_c} \mathcal{R}_{n_o,k}$$

where $\mathcal{R}_{n,k} = \prod_{i=k+1}^{n} \binom{i}{2}$, the number of ranked ways to reduce $n$ lineages to $k$.

Now it is straightforward (but tedious) to show that

$$\frac{n!}{(n-k-2)!(k+1)!} \binom{n-k-2}{n_c-2} \mathcal{R}_{k+1} \mathcal{R}_{n_c} \mathcal{R}_{N_o,k} = \frac{n_c^3 - n_c}{2} \binom{n_o-1}{k-1} \mathcal{R}_n^{(n_c)}.$$

After replacing the above and factoring out,

$$f_{T_\phi}(x) = $$
$$\frac{n_c^3 - n_c}{2} \lambda e^{-3\lambda x} (1 - e^{-\lambda x})^{n_c-2} \sum_{k=1}^{n_o} \binom{n_o-1}{k-1} (1 - e^{-\lambda x})^{n_o-k} (e^{-\lambda x})^{k-1}$$
$$= \frac{n_c^3 - n_c}{2} \lambda e^{-3\lambda x} (1 - e^{-\lambda x})^{n_c-2} (1 - e^{-\lambda x} + e^{-\lambda x})^{n_o-1}$$
$$= \frac{n_c^3 - n_c}{2} \lambda e^{-3\lambda x} (1 - e^{-\lambda x})^{n_c-2}$$

# D Conditional density of 3 nested clades with 3 taxa outside the main clade

Here we derive the marginal density for 3 nested calibration points in a $n+3$ taxon tree. The first (lowest) calibration point is on the root of an $n$-taxon monophyletic clade $\phi$, the second on the $n+1$ clade containing $\phi$ and one additional taxon, and the third is on the root of the tree, which includes the remaining 2 taxa. Let the heights of the calibration points be $x_0$, $x_1$ and $x_2$, where $x_0$ is the height of the root.

Of the $n+2$ heights in the tree, 3 are calibrated and $n-2$ are below $x_2$, which leaves just one height, $x$, outside $\phi$. This gives us 3 cases, the first where the two outside taxa coalesce before $x_2$ ($x < x_2$), the second where they coalesce between $x_1$ and $x_2$ ($x_1 \leq x < x_2$), and the third where $x_1 \leq x < x_0$. The marginal densities for the three cases are as follows:

$$f_1(x_0, x_1, x_2) = (n+3)!\lambda^3 e^{-2\lambda x_0} e^{-\lambda x_1} e^{-\lambda x_2} \frac{\left(1 - e^{-\lambda x_2}\right)^{n-1}}{(n-1)!}$$

$$f_2(x_0, x_1, x_2) = (n+3)!\lambda^3 e^{-2\lambda x_0} e^{-\lambda x_1} e^{-\lambda x_2} \frac{\left(1 - e^{-\lambda x_2}\right)^{n-2}}{(n-2)!} \left(e^{-\lambda x_2} - e^{-\lambda x_1}\right)$$

$$f_3(x_0, x_1, x_2) = (n+3)!\lambda^3 e^{-2\lambda x_0} e^{-\lambda x_1} e^{-\lambda x_2} \frac{\left(1 - e^{-\lambda x_2}\right)^{n-2}}{(n-2)!} \left(e^{-\lambda x_1} - e^{-\lambda x_0}\right).$$

For each of the possible $\mathcal{R}_n$ ways to coalesce $\phi$, there are $n-1$ ways to place $h$ between the $n-2$ heights of $\phi$ (case 1), only one way in case 2 (no other heights between $x_2$ and $x_1$), and again only one way in case 3, but here there are $\mathcal{R}_3$ ways to coalesce the three lineages between $x_1$ and the root. So the ratios of the three cases are $n-1 : 1 : 3$, which let us combine the $f_1, f_2$ and $f_3$ into the required density:

$$\begin{aligned}
f(x_0, x_1, x_2) &= \frac{1}{n+3}((n-1)f_1(x_0, x_1, x_2) + f_2(x_0, x_1, x_2) + 3f_3(x_0, x_1, x_2)) \\
&= (n-1)n(n+1)(n+2)\lambda^3 e^{-\lambda(2x_0 + x_1 + x_2)} \\
&\quad \left(1 - 3e^{-\lambda x_0} + 2e^{-\lambda x_1}\right)\left(1 - e^{-\lambda x_2}\right)^{(n-2)}.
\end{aligned}$$
(14)

# References

Drummond, A., Ho, S., Phillips, M., and Rambaut, A. (2006). Relaxed phylogenetics and dating with confidence. *PLoS Biol*, 4(5):e88.

Drummond, A. and Rambaut, A. (2007). Beast: Bayesian evolutionary analysis by sampling trees. *BMC Evol Biol*, 7:214.

Drummond, A. J., Nicholls, G. K., Rodrigo, A. G., and Solomon, W. (2002). Estimating mutation parameters, population history and genealogy simultaneously from temporally spaced sequence data. *Genetics*, 161(3):1307–20.

Drummond, A. J., Nicholls, G. K., Rodrigo, A. G., and Solomon, W. (2003). Genealogies from time-stamped sequence data. In Buck, C. E. and Millard, A. R., editors, *Tools for Constructing Chronologies: crossing disciplinary boundaries*, volume 177 of *Lecture Notes in Statistics*, chapter 7, pages 149–174. Springer.

Gernhard, T. (2008). The conditioned reconstructed process. *Journal of Theoretical Biology*, 253(4):769–778.

Griffiths, R. and Tavare, S. (1994). Sampling Theory for Neutral Alleles in a Varying Environment. *Royal Society of London Philosophical Transactions Series B*, 344:403–410.

Kingman, J. (1982). The coalescent. *Stochastic processes and their applications*, 13(3):235–248.

Phillips, M., Bennett, T., and Lee, M. (2009). Molecules, morphology, and ecology indicate a recent, amphibious ancestry for echidnas. *Proceedings of the National Academy of Sciences*, 106(40):17089–17094.

Rannala, B. and Yang, Z. (2005). Bayesian Estimation of Species Divergence Times Under a Molecular Clock Using Multiple Fossil Calibrations with Soft Bounds. *Mol Biol Evol*, 23:212–226.

Sithaldeen, R., Bishop, J., and Ackermann, R. (2009). Mitochondrial DNA analysis reveals Plio-Pleistocene diversification within the chacma baboon. *Molecular phylogenetics and evolution*, 53(3):1042–1048.

Stadler, T. (2009). On incomplete sampling under birth-death models and connections to the sampling-based coalescent. *J Theor Biol*, 261(1):58–66.

Stadler, T. (2010). Sampling-through-time in birth-death trees. *J Theor Biol*, 267(3):396–404.